# KIBS' Innovative Entrepreneurship Networks on Social Media

*José N. Franco-Riquelme\* [1], Isaac Lemus-Aguilar [1,2], Joaquín Ordieres-Meré[1]*

[1]Department of Industrial Engineering, Business Administration and Statistics, Universidad Politécnica de Madrid, Spain
[2]Department of Management, Economics, and Industrial Engineering, Politecnico di Milano, Italy

**Abstract**. *The analysis of the use of social media for innovative entrepreneurship in the context has received little attention in the literature, especially in the context of Knowledge Intensive Business Services (KIBS). Therefore, this paper focuses on bridging this gap by applying text mining and sentiment analysis techniques to identify the innovative entrepreneurship reflected by these companies in their social media. Finally, we present and analyze the results of our quantitative analysis of 23.483 posts based on eleven Spanish and Italian consultancy KIBS Twitter Usernames and Keywords using data interpretation techniques such as clustering and topic modeling. This paper suggests that there is a significant gap between the perceived potential of social media and the entrepreneurial behaviors at the social context in business-to-business (B2B) companies.*

## 1 Introduction

The use of social media for entrepreneurial innovation in the business-to-business (B2B) sector has missed specific attention (Kärkkäinen et al., 2010).

Knowledge Intensive Business Services (KIBS) are such organizations that primarily add value through the accumulation, creation or dissemination of knowledge for the purposes of the customer and which have other businesses as their main clients (Miles et al., 1995). Among KIBS, consulting firms are a major source of knowledge-based innovation because of their expertise, structure, diversification and continuous creation/recombination of knowledge (Anand et al., 2007; Wright et al., 2012).

Twitter Social Media is a prime platform for developing a framework that allows information retrieval to support analysis and management of Big Data on social media (García-Crespo et al., 2017). This professional Twitter platform can be considered part of the entrepreneurial innovation ecosystem of KIBS, in other words, they are part of the community of actors interacting with KIBS as a unique system to produce inter-organizational streams of continuous innovation and entrepreneurial behavior (Autio et al., 2014; Gastaldi et al., 2015).

To our best knowledge, only a few works address the topic of innovation in the Twittersphere (García-Crespo et al., 2017), but none of them focuses on KIBS nor the interlink between the firm's innovation management and their innovation entrepreneurial networks on social media. This study aims to contribute to filling this research gap by identifying the innovation networks reflected by these companies in their social media as well as their clustering and composition.

In this study, we perform social media analysis of followers on Twitter of eleven (11) Spanish and Italian Small and Medium Enterprises (SME´s) and large companies. We applied Big Data and opinion mining techniques to get insights from data, and we retrieve information related to each account. For this task, we have established the relationship of networks regarding keywords, e.g., innovation, entrepreneurship, etc. to carry out our exploration of terms associated with it. We have focused on quantitative approach mainly based on data analysis from social media.

This article was organized as follows. In this section, a literature review has been presented to get essential aspects of previous works. The methodology describes the sentiment analysis and machine learning techniques, performed to our data collection of tweets. Then, the results of the quantitative analysis performed through visualizations and tables are shown. A discussion regarding the results and the conclusion are presented in the last section.

## 1.1 Literature Review

### 1.1.1 Sentiment analysis on social media

Since the rise of web 2.0 we have been witnesses of the blast-off of social media in many fields. Through this way, users contribute to create their own content, distribute it, and share with others (Hanson et al., 2010), between companies, politicians, artist and ordinary citizens allowed new forms of interactions.

To assess the role of social media and its impact on the relationship between companies and users, modern online social and technological systems are producing real changes in the traditional networking paradigms and the way that we communicate each other (Borge-Holthoefer et al., 2011). In that sense, the activity of the users have been increased in the last few years through platforms like Facebook, Twitter, LinkedIn, Instagram, YouTube, among others.

In order to get relevant insights from social media, the platform chosen for this work was Twitter, which is a popular microblogging site, where users search information like breaking news, post about celebrities, and comment about users (that include companies, particulars, etc.). In addition, Twitter is an ideal source for spotting the information about societal interest and general people´s opinion (Khan et al., 2015) and for this reason; we can study the perception and interaction between users and KIBS firms and networking.

It consists of short messages called "tweets" limited to 280 characters[1] in length and can be viewed by user´s followers. In this way, researchers have been increased in the fact that people are expressing their emotions, and their likes (favorites) and retweets in some situations (Cesteros et al., 2015), giving us a particular type of measure that could be useful for the users and it network behavior.

In this research, considering Twitter has high capabilities to study the behavior of KIBS ecosystem and working with big data applications in order to get sentiment analysis, considering the large quantities of information that we can retrieve, and processed in a faster way, enabling sentiment-related insights that would be hard to determine with small data amounts (Thelwall, 2016).

In that way, opinion mining (also called sentiment analysis), is concerned with developing software to automatically retrieve opinions, about products or entities from text. Research in this area, started in the early 2000 decade since then much progress has been made (Mukherjee & Bala, 2017).

In summary, sentiment analysis is the process of computationally identifying, categorizing and treatment of opinion, sentiment and subjectivity of text, to determine whether the writer´s attitude with respect a particular topic (Pang et al., 2002; Villena Román et al., 2015).

---

[1] Since September, 26 2017, Twitter extended the text limit of a post on its service to 280 characters, from originally 140 characters: https://www.nytimes.com/2017/09/26/technology/twitter-280-characters.html

### 1.1.2 Entrepreneurial Innovation

According to Autio et al. (2014), entrepreneurial innovation focuses on radical innovation driven by co-creation, having multi-actor and -level processes to foster the development of entrepreneurial ecosystems (Autio et al., 2014).

We focus on what these authors refer as 'entry behaviors' in the context of established organization (intrapreneurship), which are performed by the employees; as well as the 'post-entry' behaviors, which are choices influencing the achievement of the previous ones such as the creation of social network ecosystems through social media.

## 2 Methods

In this section, we present the outlines of a novel approach regarding the combination of text mining and sentiment analysis techniques, performing quantitative methods of posts in Twitter social media. We carry out Natural Language Processing (NLP) and machine learning approach to get insights in the field of KIBS in the Twittersphere.

### 2.1 Data

For our data collection, we have focused on tweets about Spanish and Italian consultancy KIBS, as follows: Accenture, Altran, Beeva, Bip Italia, Codex SC, Deloitte, Gesor, KPMG, Oesia, Tecnocom and Vector ITC, that represent the companies who are object of this research, listed in Table 1.

**Table 1.** Consultancy KIBS sample tweets

| KIBS | Retrieved Tweets | Filtered Tweets |
|---|---|---|
| Accenture | 4.340 | 716 |
| Altran | 1.651 | 455 |
| Beeva | 1.350 | 235 |
| Bip Italia | 538 | 538 |
| Codex SC | 72 | 20 |
| Deloitte | 3.690 | 486 |
| Gesor | 130 | 19 |
| KPMG | 5.402 | 1.048 |
| Oesia | 477 | 54 |
| Tecnocom | 4.420 | 325 |
| Vector ITC | 413 | 81 |
| **Total** | **22.483** | **3.977** |

We retrieve a database of tweets, based on User Name keywords of each companies mentioned earlier. Data were collected from Twitter, consisting of a sample of 22.483 public tweets and for this matter, we ran the Twitter Scraper through Python language.

To our knowledge, we have defined a period of one year to retrieve tweets, starting on March 1st 2016 till February 28th 2017.

**Table 2.** Example of variables in the consultancy KIBS tweets dataset

| Variables | Data |
|---|---|
| Followers | 11257 |
| Retweets | 12 |
| Favourites | 53 |
| Text | "Blockchain Technology How banks are building a realtime global payment network AccentureSpain" |
| User Name | ESIC |
| Weighting | 20 |
| Indicator | 0.37500000 |

For this study, we have considered seven variables: followers of each count, retweets, favorites, the text of each tweet, username, weighing, that means the sum of favorites and retweets multiplied by two and finally, the Indicator that is the ratio of weighing divided the followers, can be seen in Table 2.

All the variables in this analysis are critical getting insights from tweets regarding the relationship between consultancy KIBS and the users who mentioned them.

### 2.2 Preprocessing

Data preprocessing consist, as we mentioned in the previous section in a dataset based on tweets regarding those eleven consultancy KIBS selected for this research, and after the collection of tweets, we filtered through keywords that are given in Table 3.

Table 3. Keywords used to filter tweets

| Innovation | Innovation, Innovative, innovate |
|---|---|
| Entrepreneurship | Entrepreneur, entrepreneurship, startup, opportunities, projects |
| Others | Technology, digital transformation, digitization, Internet of Things (IoT), Cognitive Systems, Big Data |

After this task, the amount of tweets were established to 3.977 tweets (see Table 1) and once having the reduced dataset, we perform the weighing of two metadata: favorites and retweets.

Hence we developed an indicator, based on the sum of likes (in Twitter jargon: Favorites) and the sum of retweets, but in this case, we made a weighing per two, in fact, that retweets have more value, considering the visibility of the reprint of tweets. In that sense, we have established a measurement in order to get the ratio of each tweet, giving us a parameter of the impact of it, in the network of KIBS.

Once having the final amount of tweets and our variables selected and established, we started the text cleaning, based on the text of each tweet, using Natural Language Processing (NLP) techniques. For this step, first, we proceed to create the corpus of annotated tweets.

In order to accomplish the text treatment, we used R language as a primary tool for processing our text mining, using the library "tm" (Text Mining), developed by (Feinerer et al., 2008) that has allowed us perform the cleanup of text, removing URLs, hashtags, numbers, emoticons, and stopwords (which consist of meaningless words, such as "the, "a", "to", etc.).

In the same way, we proceed to perform other NLP tasks: tokenization and normalization. The first it refers to word segmentation, or separation of words by space in white, and the second, giving coherence to the entire text (Jurafsky & Martin, 2009).

After all, we proceed to carry out the last preprocessing stage, before the feature selection. Hence, the text corpus should be transformed into a more straightforward form: term weighting (TF) method, which is defined to transform document´s value into terms, and resulting in a matrix of terms, that we called Term-Document Matrix (TDM). Automated tools are available to perform some or all of these steps (Ghiassi et al., 2013).

**2.3 Feature Extraction**

We started the feature of our TDM, performing as we know as Bag of Word (BoW) technique. This extraction feature, the sentences of the document are split into a set of words using the space or the punctuation characters (Psomakelis et al., 2014). These words (also called "unigrams") form a virtual bag of words, and there is no ordering or the connections between them.

**2.4 Polarity**

According to (Liu, 2010) the orientation of an opinion a feature indicates whether the opinion is positive, negative or neutral. The polarity of opinion is also known as sentiment orientation, opinion orientation, or semantic orientation.

Examples of polarity annotations of emotions are 1) Negative: anger, disgust, fear, etc.; 2) Positive: delight, joy, hope, etc.; 3) Neutral, which refers to reactive words, and there are no emotional words, like news, etc. (Liu, 2015).

For this work, we have selected the BLEL lexicon (Hu & Liu, 2004) in order to perform the relations of terms of the corpus getting the polarity based on 6.787 words, which 2.005 are positive and 4.782 are negative.

## 2.5 Models

In this section, two individual models are introduced to find out the best performance for finding the relationship between KIBS and users on Twitter. Through unsupervised machine learning methods: Clustering analysis and Topic Modeling, we want to get the interpretation of the data.

### 2.5.1 Clustering

Clustering is an essential method of data analytics, and it requires less computational cost that can be beneficial in data mining and knowledge discovery, e.g. web usage and social network analysis (Athman et al., 2015). For categorizing purposes, we performed clustering analysis of the entire dataset of tweets. In this case, this method proposed, generate agglomerative affinity between terms.

Hierarchical Dendrograms is a relatively easy approach for clustering, and its representations is a tree-like visualization and is based on frequency distance (Kwartler, 2017) of terms. This analysis is an information reduction that we applied to our dataset of tweets, after preprocessing and using the TDM created.

The objective of creating hierarchical cluster according to (Gironés Roig, 2014) consist of i) Create k groups of observations maximizing both intra-groups similarity and the intergroup difference. ii) The distance between the observations will allow groups of homogeneous but heterogeneous with each other. iii) The average of the variables that make up each cluster will help us to interpret each group. iiii) Drawing the results will determine the number of clusters to choose.

### 2.5.2 Topic Modeling

Topic models such as the Latent Dirichlet Allocation (LDA) it is a machine learning algorithm which assume a document (in our case, a set of tweets) as a mixture of topics. Since its first publication (Blei et al., 2003) it has played an essential role in a variety of text mining tasks, in fields like social and political science, bioinformatics, digital humanities, among others (Tang et al., 2014). In other words, this method could learn thematic structure from large document collections without human supervision.

In this research, we combined the extraction features related to the polarity of tweets and the topics that appeared in the dataset, resulting in an understanding broad topics, their sentiment and the number of terms devoted to the identified topic: the relationship between users of Twitter platform regarding consultancy KIBS.

## 3 Results

This section presents the results including the identification of the leading players, the quantity of post related to innovative entrepreneurship, the most popular terms used by the firms, the interrelations among the terms use the distinction of entrepreneurial behavior among consultancy KIBS.

There is a strong relation between KIBS and the social media posts from their clients for the visibility that they want to give to their events. Large corporations are the ones who have greater use of this type of social media for innovation-related diffusion. Posts also reflect the KIBS' innovation approach; however, not all consultancy firms show the same level of innovative entrepreneurship.

As a general remark, we identified that from the total bag of tweets made by KIBS in our sample; only 17.6% (3977 out of 22483) were considered innovative entrepreneurship (see Table 1). Also, only the name of 7 of the 11 consultancy KIBS in the sample are mentioned in at least one of the following analysis results, and only 5 appear in all the analysis.

Table 4 shows the top twenty-ranked KIBS influencers who tweet about entrepreneurship and innovation. 12 out of 20 influencers are employees of BIP Italia at the entry and middle management levels. Other KIBS such as Altran, Accenture and Vector ITC have one influencer belonging to the middle to top-management level.

**Table 4.** Top ten influencers in KIBS tweets network

| No. | Tweeter Account | Influence Ratio | Entity | Location |
|---|---|---|---|---|
| 1 | Stigmergy | 1 | IT News | USA |
| 2 | Mickytata | 0.875 | BIP | Italy |
| 3 | AcciónAntiBancaCGT | 0.8 | News | Spain |
| 4 | Gloria Sbrocca | 0.782 | BIP | Italy |
| 5 | Marcial Fernández | 0.636 | ICEA | Spain |
| 6 | Silvia | 0.575 | BIP | Italy |
| 7 | Paolo | 0.526 | BIP | Italy |
| 8 | filippo.pieruzzi | 0.5 | BIP | Italy |
| 9 | Alessio | 0.461 | BIP | Italy |
| 10 | Carlos | 0.406 | Altran | Spain |
| 11 | Bruno | 0.395 | BIP | Italy |
| 12 | Luis del Río | 0.384 | N/A | Spain |
| 13 | Luigi Sambuco | 0.382 | BIP | Italy |
| 14 | Ludovico Cestari | 0.375 | BIP | Italy |
| 15 | Alex Borrell | 0.307 | Accenture | Spain |
| 16 | Andrea Casati | 0.278 | BIP | Italy |
| 17 | Luca Di Traglia | 0.262 | BIP | Italy |
| 18 | Rafa Conde | 0.258 | Vector ITC | Spain |
| 19 | Umberto Petrone | 0.24 | BIP | Italy |
| 20 | Stephan Wennekes | 0.235 | Altran | Luxembourg |

In one hand, Figure 1 shows the words associated to the term 'entrepreneurship'. The words related to academia and universities are the most numerous: urjonline, laupm, ilcbarcelona, ieuniversity, universitario, actuaupm, mooc, lab, among others.

Other words are related to actors and actions is the innovative ecosytem such as government (bilbaoekintza, patrocinio, analizando, ayudando, kbidigitalgracias, impacthub, fundacion, wanajump, temprana, edad, eneste, apoyamos, transferir, momentos, apoya, impulsan, fomentan, tecnológico) or the links with a geographical region (latinoamerica, colombia).

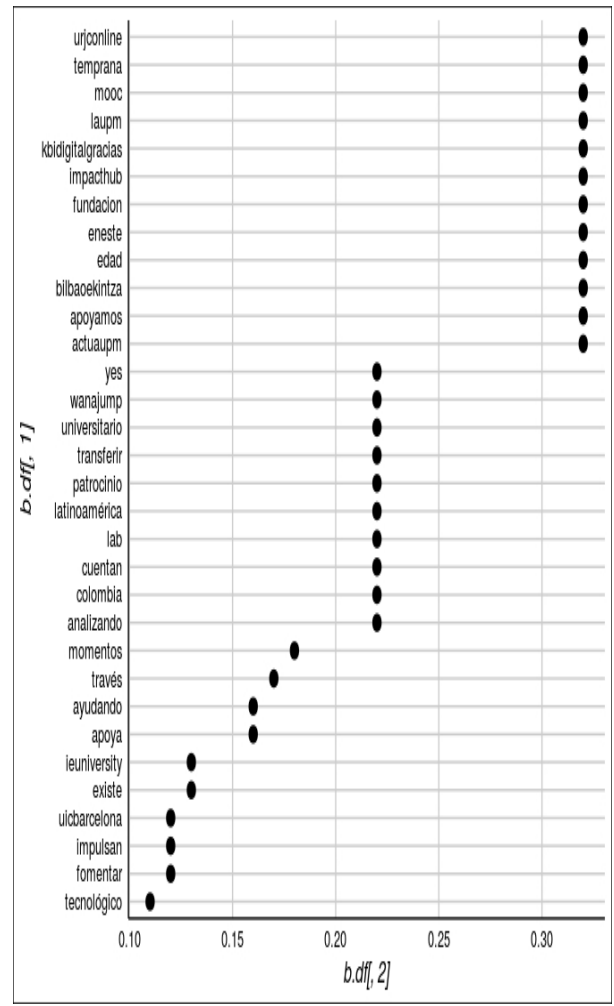

**Figure 1.** Association of words: Entrepreneurship

In the other hand, Figure 2 shows the words present in the tweets that are correlated with 'innovation', which is related to concepts such as innovation management (index, center, actitud, sostenible, nuevo, creativa, prioridad, economico, turistica, fiscal), key actors (garrigues, larracoechea, altranes, pablolinde, altran, promarcasspain) and venues (centro, ulmp, torre, picasso, innsite, traslada, abrirá).

jornada, presenta, innovative, futuro, estrategia, nuevo, big data), and key actors and places (fabiotroianibip, directivos, business, empresas, españa, sector, negocio, madrid, centro, mayor, grupo).

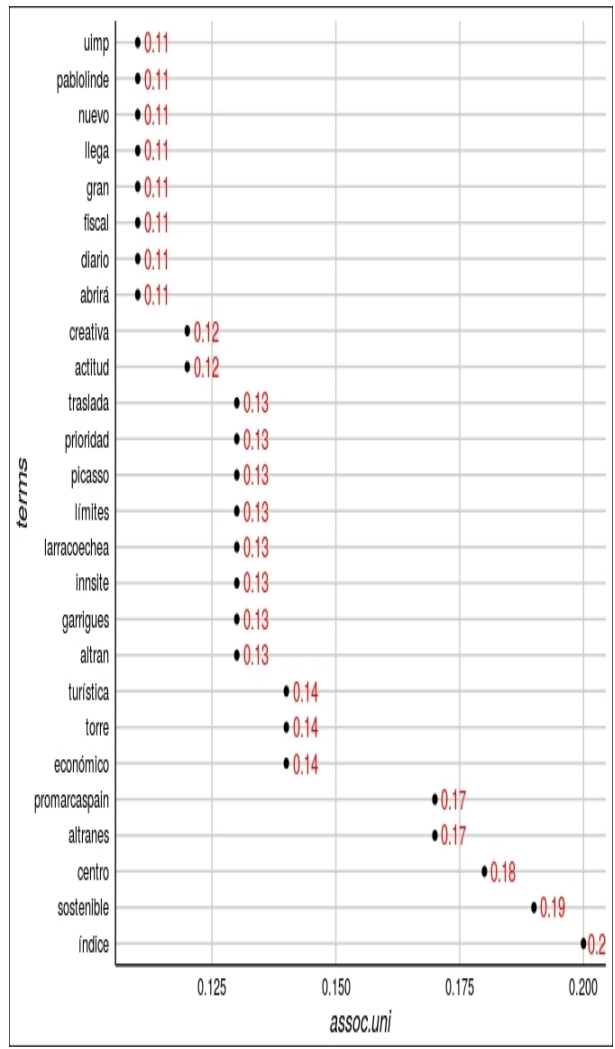

**Figure 2.** Association of words: Innovation

Figure 3 shows the most cited words in the tweet text body, which are mainly related to the KIBS Twitter accounts (bipitalia, kpmges, accenturespain, deloittees, altranes, tecnocom, beevaes).

About innovative entrepreneurship (innovacion, transformacion, tecnologia, proyectos, oportunidades, internet, talent, datos, informe, digitalizacion, openhow, nuevas, fondamentale, digitales, tendencias, tecnologicas, retos, beneficios, crecimiento,

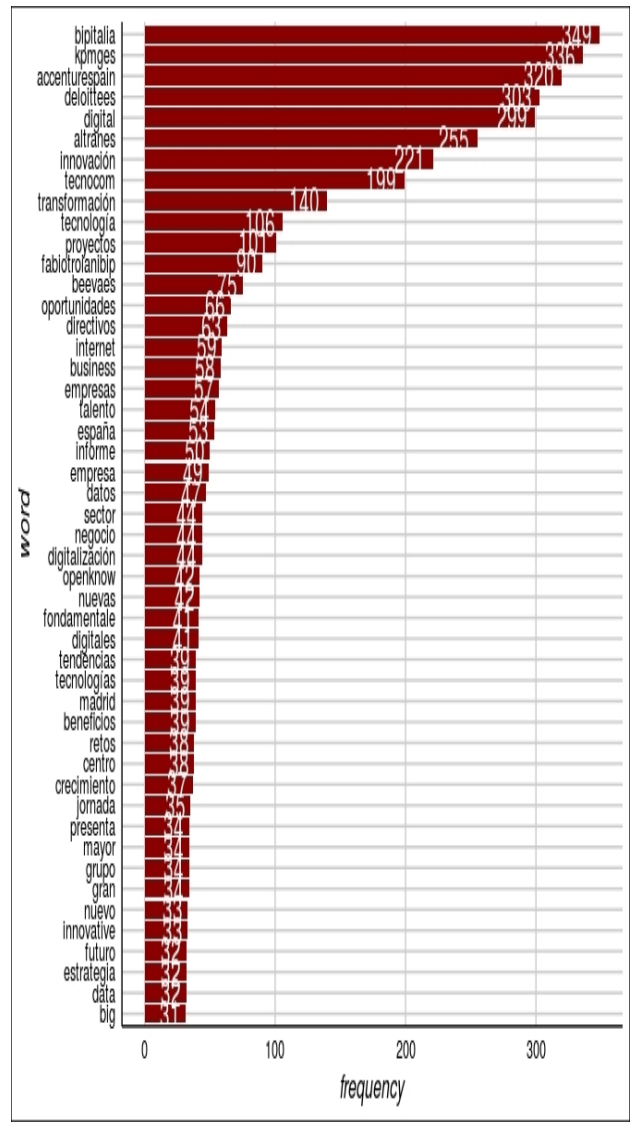

**Figure 3.** Unigrams frequency of Twitter Dataset

Taking a further step, Figure 4 gives a glimpse about what innovative entrepreneurship means for different actors involved in the entrepreneurial innovation ecosystem. For example, in the left side we can see that three companies (Tecnocom, KPMG and Accenture) focus on digital innovation and have created a social media network on this matter (transformacion, tecnocom, digital, proyectos, negocio, sector, kpmg, jornada, digitalizacion, informe, crecimiento, accenturespain, oportunidades, retos, beneficios, directivos, internet). Tecnocom mainly focus on growth by business digital transformation projects, KPMG on reporting via Internet what are the sectoral growth benefits of digitalization to top management; and finally, Accenture on the challenges and opportunities brought by growth in digitalization.

The right side of Figure 4 shows other perspectives on innovative entrepreneurship that emerged from the data. Altran focused on the creation of their innovation center in Madrid (altranes, centro, innovacion, madrid), while BIP Italia focused on the business alliance with Open Knowledge (bipitalia, business, openknow, fabiotroianibip) and Deloitte on their employees skills (deloittees, talent). Another trending topic was relate to new digital technologies (tendencias, digitales, nuevas, tecnologia).

In order to triangulate and validate the previous relation of terms, Figure 5 distinguishes the clustering of the unigram terms listed in Figure 3. The clustering results are similar to the ones in Figure 4. We can identify two main clusters; the upper one clearly shows the strong use of social media related to innovation by Altran, Deloitte and Accenture in Spain. While the lower cluster shows the focus on digital transformation by BIP Italia, KPMG, Tecnocom and BEEVA. The latter group has a broader spectrum of topics related to the ones described in Figure 3.

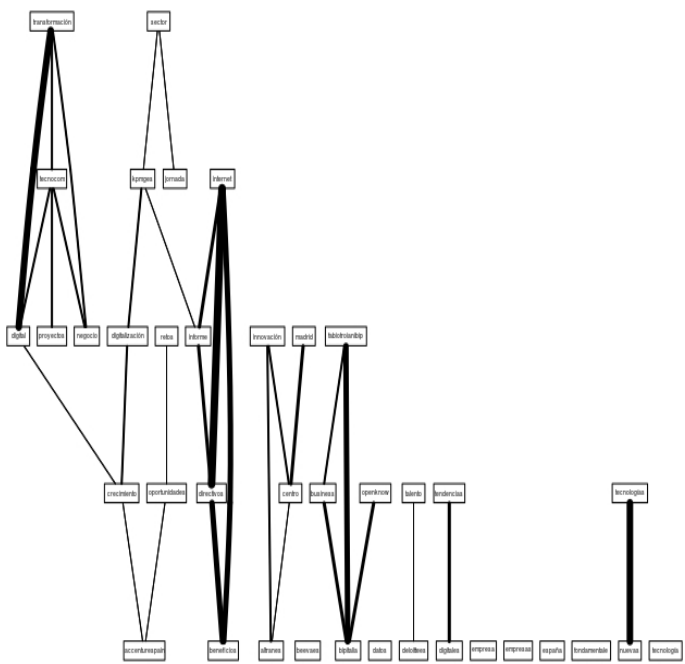

**Figure 4.** Relation of terms

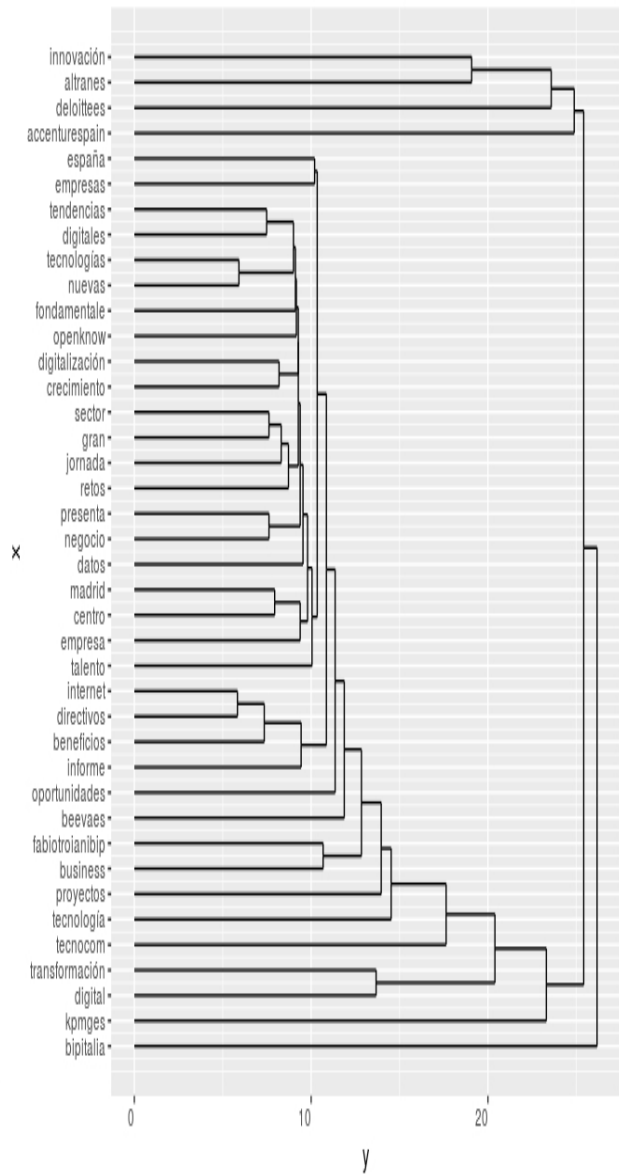

**Figure 5.** Clustering analysis

Figure 6 represents the results of the sentiment analysis performed to our collected data. The results show that the vast majority of tweets are positive in nature (green color) or neutral (light red, light green). This is no surprise, since social media is not the usual channel to solve conflicts and problems between KIBS and their clients and suppliers. Note that the sentiments is strongly positive in the upper left concepts relate to the presentation of digital transformation projects and businesses by Tecnocom, BIP Italia and KPMG Spain; as well as in the upper right corner related to the use of artificial intelligence and design as part of the consulting teams in BIP Italia and Accenture Spain.

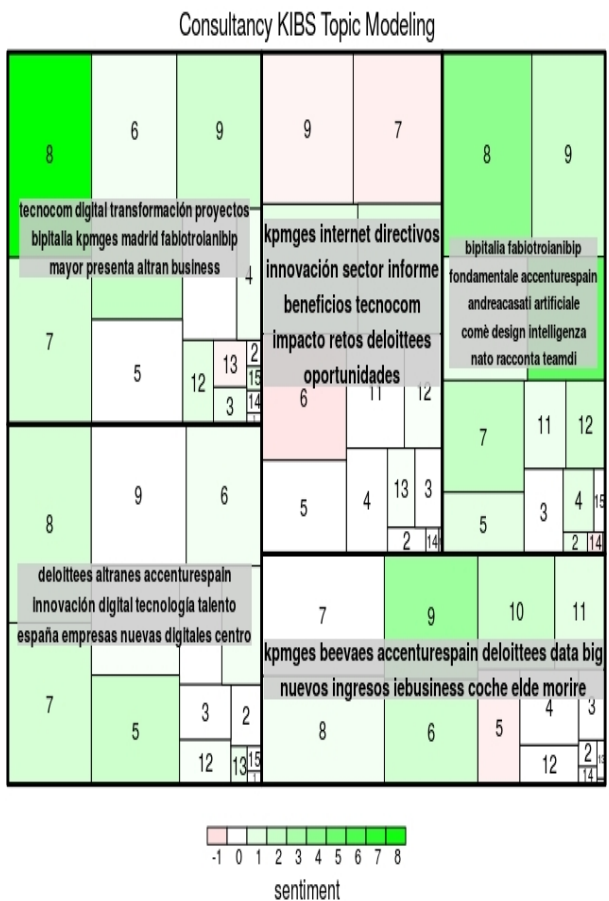

**Figure 6.** Topic modeling and polarity of tweets

## 4 Discussion

Entrepreneurial innovation involves the disruption of existing industries and the creation of new ones.

Autio et al. (2014) propose a framework of entrepreneurial innovation and context. They state that entrepreneurial behavior also happens at the social context. In our case, this social context refers to the networks between intrapreneurs and the stakeholders surrounding the KIBS firms where knowledge is shared and collaboration is set to create and discover new ideas and assets. This entrepreneurial behavior might be present in BIP Italia, whose entrepreneurial employees (intrapreneurs) account for 60% of the Top 20 influencers on this area (see Table 4).

It also needs to be distinguish that these employees are not top managers as in other KIBS such as Altran, Vector ITC or Accenture. Thus, they are mainly junior and senior consultants in BIP Italy located in different cities (Rome, Milan, Venice) that are eager to communicate what they and their firms are doing in relation to innovation and try to create this sharing knowledge ecosystem that enables them to innovate.

The evidence suggest that BIP Italia employees have "entry" and "post-entry" entrepreneurial behavior as defined by Autio et al (2014). It would be interesting to know what makes this company different from other KIBS that do not show this type of behavior.

This behavior of creating social networks are in lines with the theory of entrepreneurship proposed by (Leyden et al., 2014). This social aspect of entrepreneurship increases innovation and reduces uncertainty, and it is performed by the same actors that started the entrepreneurial action, in this case the KIBS consultants, the intrapreneurs mentioned forehead and whose names appear as influencer in the different quantitative techniques that we conducted in this study.

Our study also sheds some light into the use of digital footprints in entrepreneurship research which is one of the new theoretical gaps that need to be fulfilled (Obschonka et al., 2017). We have extended the development of an innovative entrepreneurship indicator that measures social media influencer in the consultancy KIBS ecosystem as proposed in other related studies (Low & Isserman, 2015). Finally, addressing the call by Autio et al. 2014, we have started to explore the link between entrepreneurial behaviors and social context in an innovative way using social media data and machine learning quantitative techniques.

## 5 Conclusion

This paper contributes to the understanding of entrepreneurial innovation in the context of KIBS and social media, specifically when it comes to the creation of innovation networks and ecosystems. It also sheds light on the use of big data techniques for sentiment analysis and machine learning algorithms in a specific study case that which has been little analyzed, and furthermore, it adds knowledge to the use new quantitative methods in entrepreneurial innovation studies. In conclusion, this work contributes to the use of machine learning techniques applied to social sciences. There is a limitation for reliability in the use of posts for big data analysis because people tend to post more positive comments and neglect the publication of complaints and negativity (see Figure 6).

It would be interesting to conduct new mixed methods studies having a multinational Social Media databases as well as other types of KIBS. It would be interesting to further develop an organization model to generate knowledge for entrepreneurial innovation network decision-making based on social media analysis. i.e., to estimate the relationship between social media networks and actual innovative entrepreneurship projects and alliances in the B2B arena.

## Acknowledgments

We would like to thank the support of Erasmus Mundus Action 2 (EM2) Sustain-T program. This paper is also produced as part of the EMJD Program European Doctorate in Industrial Management (EDIM) funded by the European Commission, Erasmus Mundus Action1.